\documentclass[12pt]{iopart}

\usepackage{graphics,graphicx}
\usepackage{color}

\begin{document}

\title[Drying and percolation]{Drying and percolation in spatially correlated porous media}

\author{Soumyajyoti Biswas$^{1}$, Paolo Fantinel$^1$, Oshri Borgman$^2$, Ran Holtzman$^2$ and Lucas Goehring$^{3*}$}
\address{$^1$Max Planck Institute for Dynamics and Self-Organisation (MPIDS), G\"ottingen 37077, Germany}
\address{$^2$Department of Soil and Water Sciences, The Hebrew University of Jerusalem, Israel}
\address{$^3$School of Science and Technology, Nottingham Trent University, Nottingham, NG11 8NS, UK}
\eads{$^*$\mailto{lucas.goehring@ntu.ac.uk}}

\date{\today}

\begin{abstract}
\noindent  We study how the dynamics of a drying front propagating through a porous medium are affected by small-scale correlations in material properties.   For this, we first present drying experiments in micro-fluidic micro-models of porous media.   Here, the fluid pressures develop more intermittent dynamics as local correlations are added to the structure of the pore spaces.  We also consider this problem numerically, using a model of invasion percolation with trapping, and find that there is a crossover in invasion behaviour associated with the length-scale of the disorder in the system.  The critical exponents associated with large enough events are similar to the classic invasion percolation problem, whereas the addition of a finite correlation length significantly affects the exponent values of avalanches and bursts, up to some characteristic size.   This implies that the even a weak local structure can interfere with the universality of invasion percolation phenomena.  
\end{abstract}



\maketitle

\section{Introduction}    

Intermittent dynamics, where short bursts of activity are separated by much longer quiescent (\textit{i.e.} event-free) intervals, appear in diverse physical, biological and even social systems \cite{Bak1987,Paczuski1996,Sethna2001,Kawamura2012}. Examples range from earthquakes \cite{Sornette1989} or the stick-slip behaviour of a block sliding with friction \cite{Gomes1998}, avalanches in granular materials~\cite{Denisov2016} or during fracture propagation~\cite{Tallakstad2013}, energy transfer in turbulence~\cite{Salazar2010}, the growth of reactive flows~\cite{Chevalier2017} or biological films~\cite{Pelce2004}, to variations in stock markets~\cite{Gabaix2003}, blackouts in power grids~\cite{Dobson2007}, and models of the punctuated equilibrium theory of evolution \cite{Bak1993}.   Much of the complex behaviour of these systems can be captured by surprisingly simple models.  Specifically, the statistics of intermittent dynamics are routinely found to be equivalent to invasion percolation \cite{Wilkinson1983}, a model of a network of fragile bonds which fail as a system is stressed.  In this model a failure front, corresponding to the leading edge of a crack, fluid invasion, reaction or drying front, drives its way through the network, and the activity of this front is found to obey well-defined and universal scaling laws \cite{Roux1989,maslov95}.  

We consider how local correlations in the strength of a system can modify the universal response of invasion percolation.  We are inspired in this by work in fluid-fluid invasion \cite{Ioannidis1993,Knackstedt2001,Murison2014}, and the drying of porous media \cite{Borgman2017,Fantinel17}, which have demonstrated that local correlations in the pore-scale properties of a granular medium can dramatically modify any fluid flows within it.   For the example of immiscible flows, as occur when oil displaces water, increasing the correlation length of pore-scale disorder has been shown to decrease the residual saturation of the wetting phase at breakthrough~\cite{Ioannidis1993,Knackstedt2001}, leading to a more gradually-varying capillary pressure-saturation relation~\cite{Rajaram1997, Mani1999}, and improving the connectivity, and hence permeability, of both phases~\cite{Mani1999}. Changes in fluid retention were also observed upon varying the correlation length of particle wettability in a bead pack~\cite{Murison2014}.  There is a wide range of practical applications which rely on fluid-fluid displacement, such as controlling the budget of water, carbon and nutrients in soils and rock. Some specific examples include the production of groundwater and hydrocarbons, monitoring or assessing the contamination of soils and water sources and their remediation, the safe storage of hazardous wastes and carbon sequestration~\cite{sahimi-ftpm,Bultreys2016}. 

\begin{figure}
\center
\includegraphics[width=145 mm]{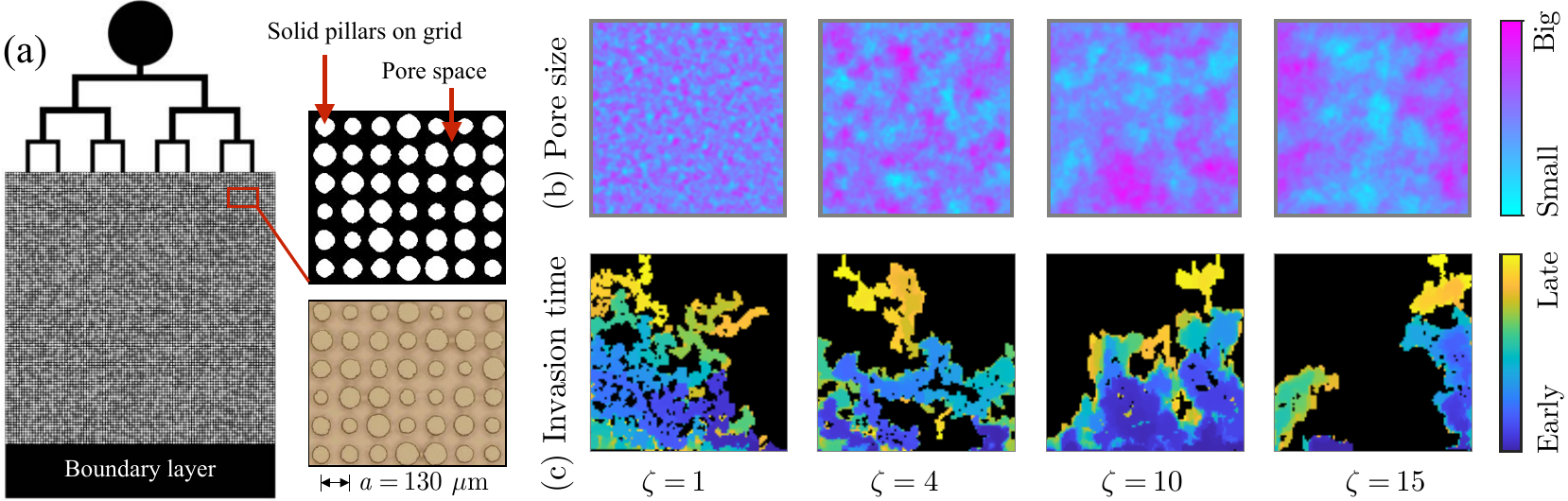}
\caption{Invasion into correlated porous media.  (a) A flat pore space is designed and fabricated in a micro-fluidic chip by soft lithography techniques.  The drying cell consists of an array of pillars, which surround pores.  The pores are initially filled through channels along one edge of the cell and then dry through the opposite edge, which is open to the atmosphere.  (b) Correlations between pillar sizes, and hence pore sizes, are randomly introduced with different correlation lengths $\zeta$.  (c) The drying pattern (shown here at breakthrough, when the invading phase connects to the back of the chip) and sequence of invaded pores reflect the underlying structure of pillar sizes.  }
\label{fig01}
\end{figure}

\begin{figure}
\center
\includegraphics[width=135 mm]{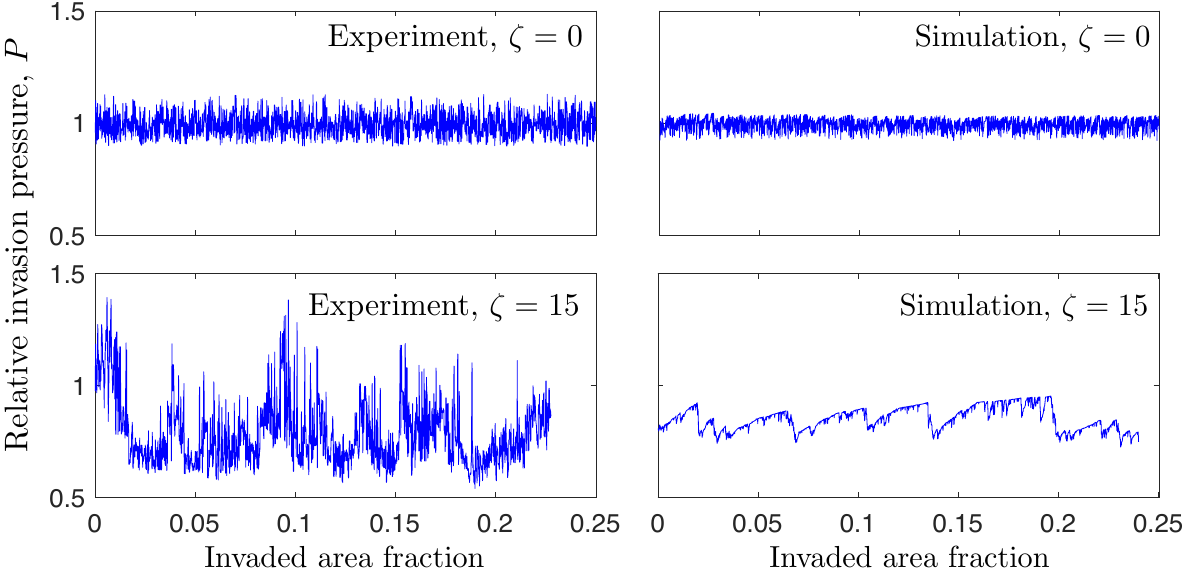}
   \caption{Local correlations in the sizes of pores and throats dramatically affect how the fluid pressure fluctuates as a porous medium is dried, in both experiments and a complimentary invasion-percolation model. When there are no correlations ($\zeta = 0$) the pressure sequence is similar to white noise, centred around the average invasion pressure of the porous medium ($P=1$).  In contrast, when the correlation length is $\zeta = 15$ times the typical pore size, the pressure sequence shows strong intermittency.}
\label{fig02}
\end{figure}

Here, we demonstrate that spatial correlations in the pore sizes of a porous medium affect its intermittent dynamics, \textit{via} the particular case of drying.  During drying, evaporation of a defending fluid increases the curvature of any menisci caught in the pore spaces, resulting in the capillary invasion of air, once a local pressure threshold is exceeded.  We study, both experimentally and numerically, the invasion patterns and statistics of a drying front moving through a disordered porous medium, where the sizes of nearby pores are correlated with each other.   The experimental realisation is a quasi-two-dimensional micro-fluidic chip, where fluids can move around an array of rigid pillars arranged on a regular grid \cite{Fantinel17}, as described in Fig. \ref{fig01}.  The pillar radii are correlated over a given length scale $\zeta$ \cite{Borgman2017,Fantinel17}.  The numerical simulation is that of invasion percolation on a grid, with its parameters taken from the geometry of the experiments.   In both cases, as shown in Fig. \ref{fig02}, the capillary pressure at the drying front fluctuates, and spatial correlations change the statistics of these fluctuations, such as the likelihoods of extreme pressures, or the distributions of sudden bursts of activity.  

Our primary motivation is to explore and explain the effects of such a correlation length, intermediate between the pore or grid-scale and the system size, on invasion percolation.  Given that the intermittent dynamics of the drying front is a critical phenomenon, its statistics can be characterised by a set of critical exponents \cite{martys91a, moura17,moura17b, aker00}.  We find that these exponents are changed by the introduction of the correlation length scale. Furthermore, returning to the experimental motivation, we investigate these effects and determine the scaling relations for drying interfaces in correlated media.

\section{Methods}

\subsection{Microfluidic experiments}

Experiments were conducted in micro-fluidic chips containing a mechanical micro-model of a porous medium; detailed methods of their design and fabrication are given in Refs. \cite{Fantinel17,Borgman2017}.  Briefly, we used soft lithography techniques to create a pattern of 100 $\times$ 100 solid pillars on a regular grid, of spacing $130$ $\mu$m, defining a pore space between them.  The pillars had an average radius of $a = 50$ $\mu$m and height $h = 40$ $\mu$m and the pore space was designed in a Hele-Shaw geometry, as in Fig. \ref{fig01}(a).  To introduce disorder, the pillar sizes were randomised over the range of 40-60 $\mu$m.   For samples with no correlations, the size of each pillar was selected randomly from a uniform distribution over this range.  To make geometries with correlations in pillar sizes, such as patches of larger or smaller pillars, we generated random Gaussian rough surfaces \cite{Borgman2017,Persson2005} with a characteristic correlation length $\zeta$.  Pillar sizes were then assigned by sampling this surface at the grid positions, and mapping the resulting values onto the range of 40-60 $\mu$m. The manufacturing tolerance on these pillar sizes is estimated at about $\pm$3\% \cite{Fantinel17}. The length $\zeta$ defines the distance, measured in grid spacings, over which nearby pillars will have roughly similar sizes.   Example designs are shown in Fig.~\ref{fig01}(b).

The system was then filled with a wetting volatile oil and allowed to dry from one edge of the cell. The drying pattern was imaged by an overhead camera, and the time-lapse images were used to find the time at which any given pore was invaded by air, as in Fig. \ref{fig01}(c).  By identifying the widest exposed throat on each invaded pore, at the moment of its invasion, the sequence of pore invasions was used to determine the sequence of invasion pressures, during drying. We consider the absolute value of the pressure sequence, for simplicity, although the actual capillary pressure will be negative (\textit{i.e.} lower than atmospheric).  These pressures were then normalised by the characteristic pressure $\bar{P} = 2\gamma(1/w + 1/h)$, where $w = 30$ $\mu$m is the average throat width, and $\gamma$ is the surface tension.  In this study we focussed on the relative pressure, $P$, of the main cluster, which is the set of filled pores that remain connected to the rear of the chip.  As shown in Fig.~\ref{fig02}, the observed pressure fluctuations of the main cluster show intermittent dynamics, which depend on the range of correlations in the experiment.  

\subsection{Invasion percolation model}

\begin{figure}
\center
\includegraphics[width=145 mm]{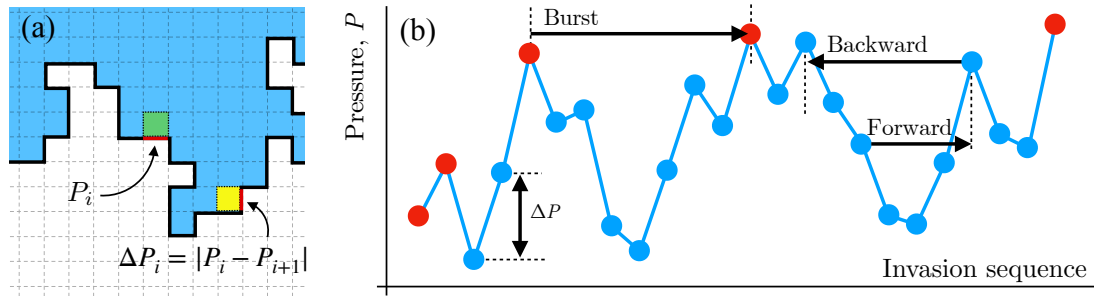}
\caption{Model and statistics of the pressure-saturation fluctuations.  (a) The invasion-percolation model tracks an interfacial line between wet (blue) and unsaturated (white) regions.  At every time step $i$ the interfacial throat with the lowest absolute invasion pressure $P_i$ is broken, and air invades the pore behind it.  Two example invasions are shown, highlighted in yellow and green. The difference $\Delta P$ measures the absolute value of the change in the fluid pressure between the successive invasion of two pores.   (b) Bursts are defined by record-breaking values of the pressure (here, highlighted in red).  Their size is measured by the number of subsequent events required until the previous maximum pressure is exceeded.  Forward and backward avalanches are defined, in contrast, for \textit{every} invasion step (as in \cite{Roux1989,maslov95}).  Their sizes, $S_f$ and $S_b$, give the number of invasion events required before that moment's invasion pressure is next exceeded, or was last seen, respectively. }
\label{fig03}
\end{figure}

In order to simulate these dynamics we used a minimal model of the drying front -- \textit{i.e.} the fluid-air interface of the main cluster.  The pore throats each have a threshold pressure (again, taking the absolute value, relative to atmospheric pressure), which must be exceeded in order to invade a pore through that throat.   As in the experiments, we considered a square lattice of pillars, using the same algorithms for randomly selecting the pillar sizes, with correlation length $\zeta$.   The separations between adjacent pillars give the throat apertures, and hence the invasion pressures. A simulation begins with a saturated pore space, and is invaded by air from one edge. The weakest throat exposed to the invasion line, or drying front, is invaded and the position of the front is adjusted accordingly.   The weakest throat along the new front is then identified, the pore behind it invaded, and so on.  An example of this algorithm is shown in Fig. \ref{fig03}(a).

The sequence of the throat invasions provides a sequence of invasion pressures, which can then be compared with experiments of similar, or identical, sample geometry.  In this study, we consider the effect of spatial correlations on the dynamics of this pressure signal, $P$. Our quasi-static model is equivalent to the invasion-percolation problem with trapping \cite{wilkinson83,Roux1989,Maloy1992,Prat1993}.  This means that the further evolution of any isolated clusters, trapped behind the drying front, are not considered.  Furthermore, the dynamics are tracked only by the sequence of pore invasions, rather than any more physically meaningful sense of time.   In these ways it differs from a related pore-network model, which has been developed to explore the impact of spatial correlations on drying \textit{rates}  \cite{Borgman2017}.  The percolation model that we consider here exactly reproduces the sequence of pore invasions along the drying front of that pore-network model, but highlights the universal aspects of the system response.  

\subsection{Burst statistics}

We use a variety of statistics to characterise the fluctuations in the system pressure, $P_i$, over the sequence $i$ of pore invasions, as shown in Fig. \ref{fig03}(b).  The power spectrum can be calculated directly from the pressure series, for a relative frequency $q$ normalised by the Nyquist or sampling rate, here that of the pore invasion rate.  Between the invasion of any two successive pores there is a change in the invasion pressure of $\Delta P_i=|P_i-P_{i+1}|$, and the intermittency of the pressure signal can be characterised by the probability distribution of these steps, $D(\Delta P)$.

\begin{figure}
\center
\includegraphics[width=145 mm]{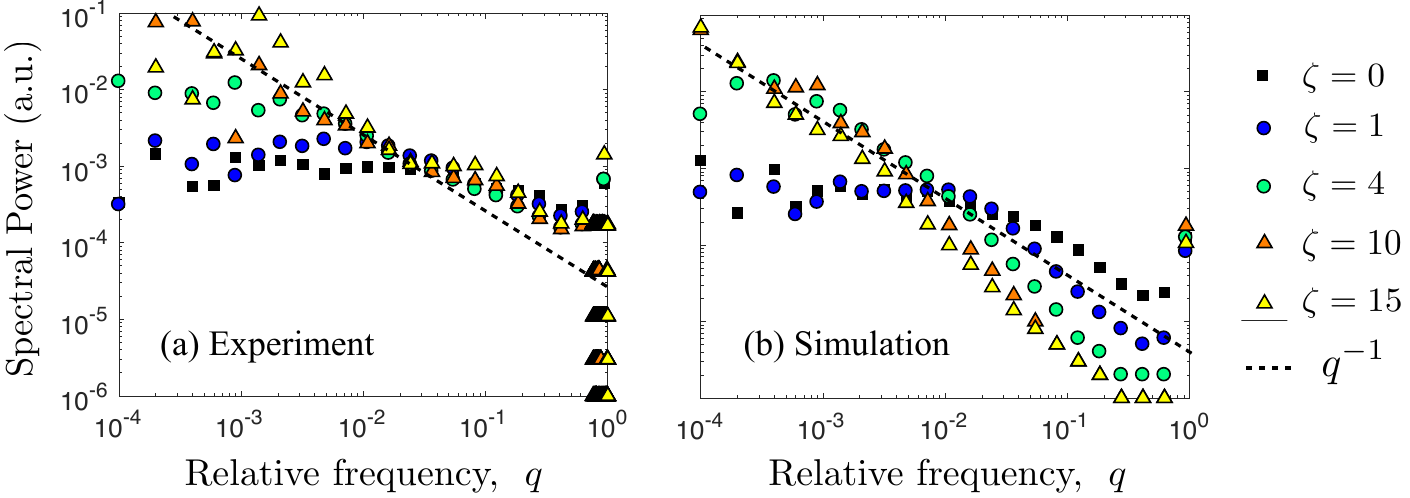}
\caption{The power spectra of the pressure signals from (a) experiments and (b) simulations respond to local correlations.  For low frequencies, in particular representing fluctuations over tens or more invasion events, the uncorrelated porous media show relatively flat power spectra, characteristic of white noise.   The introduction of correlations in the pore geometry adds significantly to the low-frequency power, which corresponds to the jumps of the pressure shown in Fig. \ref{fig02}.  For comparison, the dashed lines show a 1/$q$ response, which is classically associated with intermittency \cite{Bak1987}.}
\label{fig04}
\end{figure}

We also measure the burst size distribution $N(n)$, where $n$ is the size of a burst, defined as the number of events from one
extreme value of the pressure, until this pressure is next exceeded.  In other words, $n_i$ is the number of pores invaded between sequential record-breaking values of the fluid pressure.  For a fluid invasion experiment with a controlled injection pressure, invasion would be unstable between these points, and occur in a rapid `burst', as in Refs. \cite{moura17,moura17b}.  In contrast, for drying \cite{Prat1993,xu08} or rate-controlled fluid drainage \cite{Maloy1992} the pore pressure relaxes during pore invasion, and the intermediate pressures can be more readily observed.   

The number of record-breaking events in any particular experiment is relatively small.  Therefore, to characterise the `roughness' of the pressure fluctuations we also measure the size distributions of forward and backward avalanches, as in Refs. \cite{Roux1989,maslov95,Paczuski1996}.    After each and every pore invasion the pressure signal may take some time to recover to its former level, providing a local definition for the size of an invasion avalanche.  The size of a forward avalanche, $S_f$, is thus defined as the number of pore invasion events required until the pressure at the root of the avalanche is next exceeded.  Similarly, the size of a backward avalanche, $S_b$, is calculated for every pore invasion, measuring the number of steps that have passed since that pressure was \textit{last} exceeded.  Note that, for discrete data, these definitions are not necessarily equivalent \cite{maslov95}. 

In the following results we will show how these metrics, and their distributions, are affected by correlated disorder in both experiments and simulations of invasion percolation.  

\begin{figure}
\centering 
\includegraphics[width=145 mm]{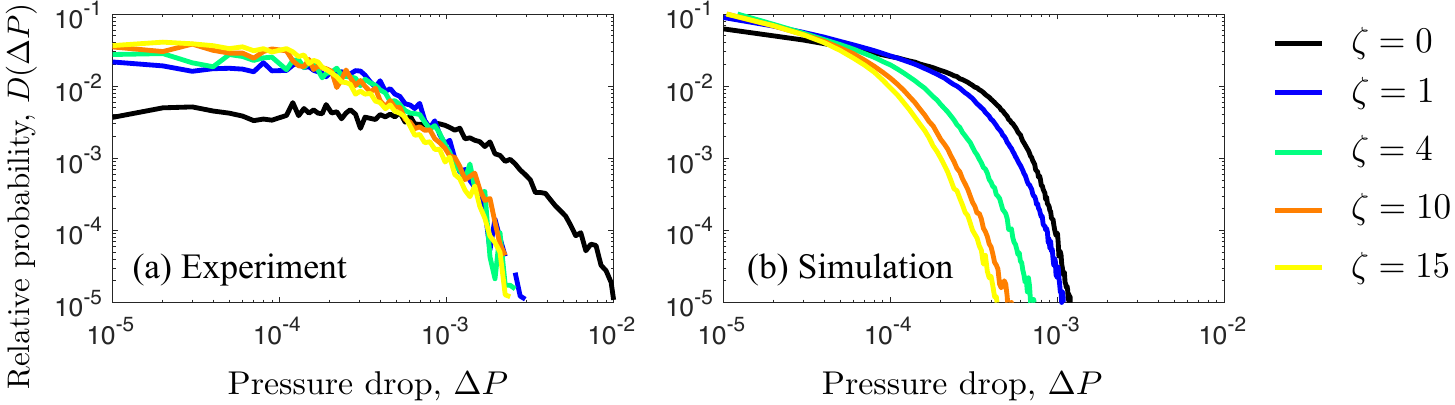}
\caption{Distributions of the pressure changes between subsequent pore invasions are shown for (a) drying experiments and (b) simulations.  As the correlation length $\zeta$ increases, there is a tendency to have fewer extreme events, as more invasion events will occur within correlated regions of similar pore size.  This is clearer in the simulation data, as uncertainties in exactly determining the experimental pressure signal introduce high-frequency random noise.}
\label{fig05}
\end{figure}

\section{Results}     

The behaviours of fluids drying in correlated and uncorrelated random media are markedly different, as was shown in Fig. \ref{fig02}.  These differences can be seen more clearly, especially in light of any experimental noise, in the power spectra of the pressure signals.  As shown in Fig. \ref{fig04}, for uncorrelated ($\zeta = 0$) disorder, the low-frequency fluctuations in the pressure signal are comparable to white noise, for both experiment and simulation.  As correlations are added, they add power to the low-frequency signal, increasing it until it approaches a $1/q$ noise spectrum.    Our aim here is to explain these differences, in light of the crossover length-scale provided by the correlation length of the disorder.

Correlations also affect the distribution of the pressure jumps between sequential invasion events, as represented by the high-frequency response of the invasion pressure. In Fig. \ref{fig05} the probability distributions for the sizes of these jumps are plotted for both experiments and simulations. The plots have similar trends, depending on the correlation length $\zeta$. In particular, an increasing $\zeta$ implies that larger jumps become less common.  This can be understood by noticing that once a pore has been invaded, it is likely to allow the drying front to explore an entire region made up of similarly-sized pores, with similar invasion pressures.  The larger, rarer, pressure jumps will frequently represent the front travelling between correlated regions, separated by areas with tighter pores.  Crossing such a high-pressure barrier triggers an avalanche, and then the pressure will start to build up again as the front explores its new patch of available pores.  Since the number of pores in a correlated region scales with $\zeta^2$, this sets a natural `timescale' for considering crossover behaviour.  

\begin{figure}
\centering 
\includegraphics[width=100 mm]{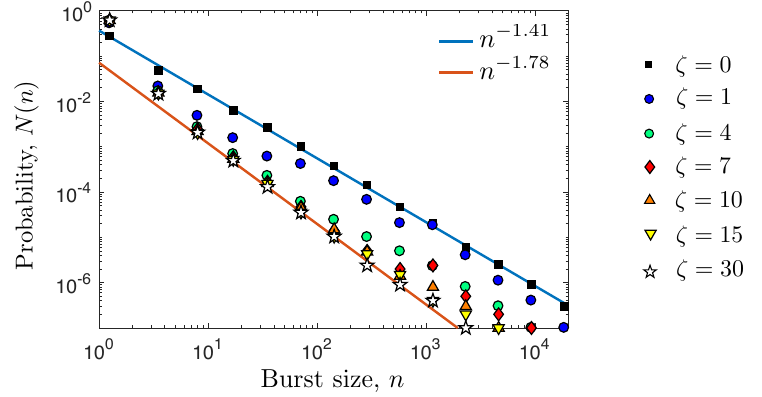}
\caption{A burst covers the invaded area between any two sequential record-breaking values of the saturation pressure.  For pressure-controlled invasion, as in Refs. \cite{moura17,moura17b}, this would be a rapid `burst' of activity between stable pressure values, while for our volume-controlled invasion it can be measured from the invasion pressure series, as in Fig. \ref{fig03}.  In the absence of any correlations in pore size, the bursts follow a power-law distribution with an exponent that is similar to values reported for other examples of invasion percolation \cite{moura17,Furuberg1996,martys91b}.  When correlations are introduced, the distribution shifts to a steeper power law, representing more frequent, but smaller, burst events, such as those confined to a single correlated patch of pores.}
\label{fig06}
\end{figure}

A common measurable of intermittent phenomena is the burst size distribution.   As mentioned in the previous section, the number of bursts in any given experimental run is rather limited, as the number of record-breaking events typically grows logarithmically in time.  Hence, we consider here only the simulation results, averaged over 60 independent realisations.  To improve the statistics for this result we also simulate larger areas, of 700 $\times$ 700 pores, and correlation lengths of up to $\zeta = 30$.  The results are presented in Fig. \ref{fig06}.  Bursts in the uncorrelated ($\zeta = 0$) porous medium show a clear power-law behaviour, with $N\sim n^{-\tau}$.  A least-squares fit to the logarithmically binned data gives an exponent of $\tau = 1.41 \pm 0.03$.  This agrees with recent experimental observations of comparable burst size statistics in drainage \cite{moura17}, where the exponent is measured at $1.37 \pm 0.08$.   It is also close to, albeit slightly higher than, the burst size distribution predicted by percolation theory in an infinite 2D system \cite{martys91b}, namely $17/13 \simeq 1.31$, and corresponding simulations \cite{Furuberg1996}.   However, as the correlation length $\zeta$ increases, the results shift away from this distribution, especially for smaller bursts.  By $\zeta = 30$ a large range of data is seen to be converging to a new power-law distribution, with the steeper exponent of $1.78\pm0.08$. For the intermediate correlation lengths, the events of smaller sizes resemble the highly correlated regime, while the larger events approach the response of the uncorrelated system.  This crossover is rather intuitive, as the smaller events preferentially sample nearby pores with similar sizes, while larger events will be limited by pores outside the immediate correlated patch, which have more random fluctuations. 

\begin{figure}
\centering 
\includegraphics[width=145 mm]{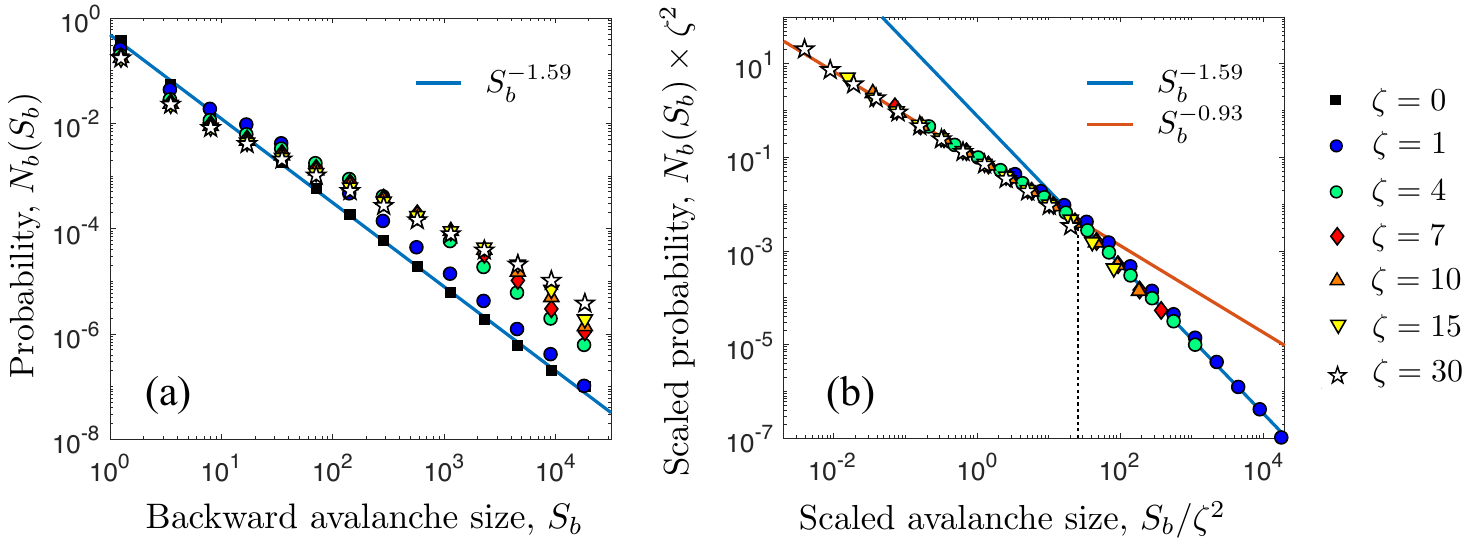}
\caption{Backward avalanches measure the number of events, going backwards in time, until the pressure of any particular invasion was last exceeded (see Fig. \ref{fig03}).  The simulated distributions (a) collapse onto a master curve (b) when scaled by the area of a correlated patch of pores, $\zeta^2$.  The crossover between these two scaling regimes, marked by a dashed line, occurs at around 5 times the correlation length, $\zeta$.  Essentially, while small bursts are strongly affected by local correlations, events large enough to average over many such correlated patches show a mean-field distribution.} 
\label{fig07}
\end{figure}

Avalanche size distributions, as defined in Fig.~\ref{fig03}(b), offer a similar metric to bursts, but allow for more statistics to be gathered.  They also show a crossover between two types of response.  We consider the size of backward avalanches in Fig.~\ref{fig07}, again for simulations of 700 $\times$ 700 pores.   For uncorrelated porous media, as in Fig.~\ref{fig07}(a), the probability distribution of the backward avalanche sizes is a power-law, $N_b \sim S_b^{-\theta_b}$, with a critical exponent fit to $\theta_b = 1.59\pm0.03$.  As expected by the theory of invasion percolation \cite{maslov95}, this value is related to the exponent of the burst size distribution by 
\begin{equation}
\label{eq1}
\theta_b=3-\tau.
\end{equation}
For invasion into correlated porous media the avalanche size distribution also changes, and becomes shallower.  When scaled by $\zeta^2$, as in Fig.~\ref{fig07}(b), the results for all correlation lengths collapse onto one master curve with two distinct scaling regimes.  Events that are significantly larger than the correlated area are consistent with the scaling of invasion percolation in uncorrelated porous media, and are fit by the exponent $\theta_b = 1.59\pm 0.06$.  Smaller events, which are more likely to explore only a space confined to a single patch of similarly-sized pores, show a power-law distribution that is well-fit by the smaller exponent $0.93 \pm 0.03$.  Interestingly, the relationship in Eq. \ref{eq1} no longer seems to hold for these local events, where $\theta_b + \tau = 2.71\pm 0.09$.  The crossover scale, estimated by the intersection of the two asymptotic curves, is about $5\zeta$ (specifically, $\zeta^2 = 23.5$). This implies that avalanche events must cover a remarkably large area before they can average over enough local structure to return to a mean-field distribution.   

\begin{figure}
\centering 
\includegraphics[width=145 mm]{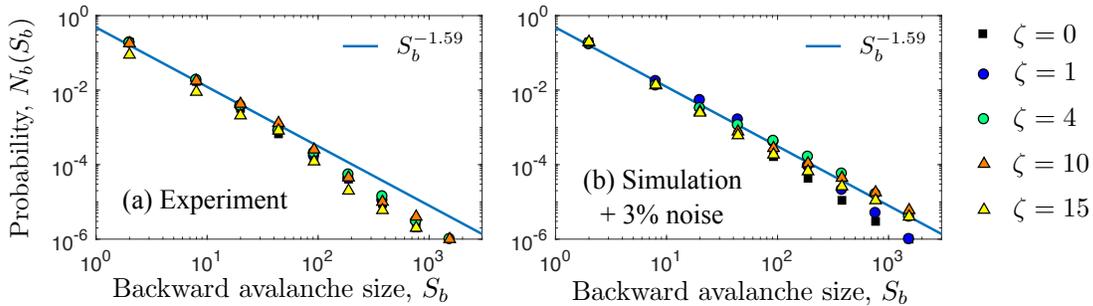}
\caption{Backward avalanches were measured in the experimental pressure signals.  (a)  There is no clear effect of correlation length in the avalanche statistics, despite its leading to intermittency in the original signal (Fig. \ref{fig02}) and power spectra (Fig. \ref{fig04}). We find, however, that the measured avalanche size distribution is strongly affected by noise, or uncertainties in the invasion pressure of each throat.  (b) Adding 3\% random noise to each measurement, comparable to our experimental manufacturing tolerance \cite{Fantinel17}, is enough to interfere with seeing the crossover behaviour of our simulations.   As a guide to the eye, curves show the fit from Fig. \ref{fig07}(a) for the noise-free $\zeta =0$ simulations.}
\label{fig08}
\end{figure}

We also measured backward avalanche sizes in the experimental invasion pressures, the distribution of which is given in Fig. \ref{fig08}(a).  The data generally follow a power-law, which is consistent with the exponent of $\theta_b = 1.59$ seen in the simulations of uncorrelated media.  However, there is no noticeable effect of correlations on the experimental results, and specifically there is no clear crossover behaviour with $\zeta$.   A possible reason for this discrepancy lies in experimental noise: there is a random manufacturing error, of about $\pm 3\%$, in the pillar radii \cite{Fantinel17}.  Since we rely on these radii to infer the pressure signal during drying, there will be a corresponding uncertainty in each pressure measurement.  To test the implications of this, we added a 3\% random perturbation to each value of the simulated pressure signals (using a 100 $\times$ 100 grid of pores, to match the experimental scale), and reanalysed the resulting data.  As can be seen in Fig. \ref{fig08}(b), even this small uncertainty is sufficient to obscure the crossover between the two limiting types of distribution.   In fact, the data from the simulations with noise now strongly resemble the experimental distributions.  

\begin{figure}
\centering 
\includegraphics[width=145 mm]{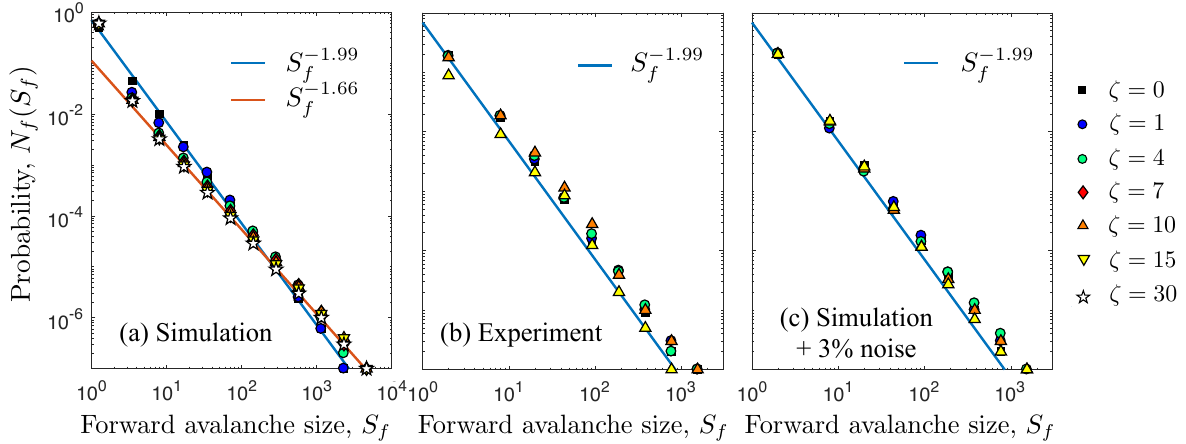}
   \caption{Forward avalanches measure the number of events, going forward in time, until any particular invasion pressure is next exceeded (see Fig. \ref{fig03}).  (a) In simulations the distribution for an uncorrelated porous medium follows a power-law distribution with exponent $\theta_f = 1.99\pm0.05$.  This is consistent with a `superuniversal' distribution, where $\theta_f = 2$, seen in a large class of invasion models \cite{maslov95,Paczuski1996}.  The local correlations modify this distribution, lowering the exponent to $\theta_f = 1.66\pm0.04$.  (b) The forward avalanches in the experiment are show little variation with $\zeta$, and are consistent with $\theta_f =2$.  (c) As with backward avalanches, the introduction of as little as 3\% noise into the pressure signal interferes with the clear observation of the crossover behaviour.}
\label{fig09}
\end{figure}

Finally, we measured the forward avalanche sizes, and their distributions, in both simulations and experiments.  The results, plotted in Fig. \ref{fig09}, show similar trends to those of backward avalanches.  For simulations of invasion in uncorrelated porous media, and the larger grid of 700 $\times$ 700 pores, the probability distribution of forward avalanche sizes follows a power law with exponent $\theta_f = 1.99 \pm 0.05$, as in Fig. \ref{fig09}(a).   Correlated disorder modifies this exponent: for the $\zeta = 30$ simulations, the data are instead well-fit by the lower exponent $\theta_f = 1.66 \pm 0.04$, for example.  However, again, this reduction in exponent cannot be detected either in the experimental data of Fig. \ref{fig09}(b), or in the (100 $\times$ 100 pore) simulations to which noise has been added to better mimic the experimental pressure signal, as shown in Fig. \ref{fig09}(c).    In a large class of avalanche statistics, or intermittent phenomena, it is known that forward avalanches follow a `superuniversal' distribution \cite{maslov95,Paczuski1996}, with $\theta_f=2$, and our $\zeta = 0$ results are consistent with that expectation.  Interestingly, the local correlations in the geometry on which invasion happens are able to modify this critical exponent, with a crossover to a shallower exponent describing avalanches that are comparable or smaller to the area of a patch of similarly-sized pores.

\section{Discussions and conclusions}

Inspired by experimental observations of drying \cite{Borgman2017,Fantinel17}, fluid-fluid invasion \cite{Ioannidis1993,Knackstedt2001,Murison2014} and drainage \cite{moura17,moura17b}, we have investigated how local correlations in the structure of an otherwise random porous medium affect the problem of invasion percolation.  For this we used a micro-fluidic micro-model of drying, which allows matched numerical modelling, by either a pore-network model \cite{Borgman2017}, or an invasion-percolation model of the leading drying front.    In all cases, the introduction of even relatively weak local correlations in the size of nearby pores and throats visibly changed the sequence of invasion pressures.  As represented by both time series (Fig. \ref{fig02}) and power spectra (Fig. \ref{fig04}), the local correlations increase the variability of the pressure signal.  Specifically, they add to the low-frequency aspects of the pressure fluctuations, leading to a pattern similar to the 1/$f$ noise traditionally associated with intermittency.  

These changes are due to the introduction of a length-scale, intermediate to the size of a single invaded pore, and the system size.  It describes the size of a patch of similarly-sized pores, and corresponds to an invasion landscape consisting of easy-to-invade `valleys', separated by more challenging `peaks'.  Such a geometry produces fewer extreme events, but adds longer-term fluctuations as activity shifts from one correlated region to another.  These effects can be seen throughout our analysis, from the power spectra of Fig. \ref{fig04}, the distribution of pressure changes between sequential invasion events (Fig. \ref{fig05}), and the probability distribution of the sizes of pressure bursts (Fig. \ref{fig06}).   Although it has been known that correlations in pore geometry can affect the breakthrough saturation and fractal dimension of invasion phenomena \cite{Ioannidis1993,Knackstedt1998,Knackstedt2001}, we have shown that they can also interfere with the expected relationships of the critical exponents of invasion problems.  For example, when correlations are significant, the critical exponents of bursts and backward avalanches obtained in our model no longer sum to 3, while that of forward avalanches is no longer required to be precisely 2.  

Surprisingly, very little correlation is required to modify the scaling laws of invasion percolation.  We found a crossover between two sets of the critical exponents that characterise the burst size distribution (Fig. \ref{fig06}), and the distributions of backward (Fig. \ref{fig07}) and forward (Fig. \ref{fig09}) avalanches.  Our geometry is constructed around an autocorrelation function $a = \exp(-\chi/\zeta)$, for two points separated by a distance $\chi$.  Here, mean-field behaviour was recovered only for events spanning an equivalent scale of $\chi\geq 5\zeta$, over which distance correlations will be virtually unmeasurable.  For flows in real systems, such local correlations in material properties can be introduced in many ways.  For example, by the roughness typical of a fracture surface \cite{scholz1995,Auradou2001}, or by a soil containing grains with a mixture of wettabilities \cite{Murison2014,Bergstad2016}, or by pore-scale correlations that are present in even relatively uniform sedimentary rock like Berea sandstone \cite{Knackstedt1998}.   One may expect, therefore, that the types of effects discussed here will apply to just as wide a range of situations.  They also highlight the danger of coarse-graining, or continuum approximations, made without taking into account disorder, and fine-scale structure, in relation to percolation phenomena.  

\section*{References}

\providecommand{\newblock}{}

\end{document}